\documentclass{aa}  
\usepackage{graphicx}
\usepackage{txfonts}
\newcommand{\be}{\begin{equation}}    
\newcommand{\ee}{\end{equation}}
\newcommand{\ba}{\begin{eqnarray}}
\newcommand{\ea}{\end{eqnarray}}
\begin{document}
   \title{Periodic orbits in the logarithmic potential}


   \author{Giuseppe Pucacco
          \inst{1}\fnmsep\thanks{also at: INFN, Sezione Roma Tor Vergata}
          Dino Boccaletti\inst{2}
          \and
          Cinzia Belmonte\inst{3}
          }

   \offprints{G. Pucacco}

   \institute{Physics Department, University of Rome ``Tor Vergata", Via della Ricerca Scientifica, 1 -- I00133 Rome\\
              \email{pucacco@roma2.infn.it}
              \and
           Mathematics Department, University of Rome ``La Sapienza", P.le A. Moro, 2 -- I00185 Rome\\  
           \email{boccaletti@uniroma1.it} 
         \and
             Physics Department, University of Rome ``La Sapienza", P.le A. Moro, 2 -- I00185 Rome\\ 
             \email{belmonte@roma1.infn.it}
             }

   \date{Received April 22, 2008; revised July 15, 2008; accepted July 16, 2008}

 
  \abstract
   {Analytic methods to investigate periodic orbits in galactic potentials.}
   {To evaluate the quality of the approximation of periodic orbits in the logarithmic potential constructed using perturbation theory based on Hamiltonian normal forms.}
   {The solutions of the equations of motion corresponding to periodic orbits are obtained as series expansions computed by inverting the normalizing canonical transformation. To improve the convergence of the series a resummation based on a continued fraction may be performed. This method is analogous to that looking for approximate rational solutions (Prendergast method).}
   {It is shown that with a normal form truncated at the lowest order incorporating the relevant resonance it is possible to construct quite accurate solutions both for normal modes and periodic orbits in general position.}
   {}


   \maketitle
%

\section{Introduction}

   In his book on Dynamical Astronomy, Contopoulos (\cite{c1}) encourages to investigate higher-order versions of the Prendergast (\cite{prende}) method to solve non-linear differential equations. The original method was applied by Contopoulos \& Seimenis (\cite{contos}, hereafter CS90) to periodic orbits in the logarithmic potential and consists in approximating the exact solution with rational trigonometric functions. Even though the trigonometric series used in the rational approximation are truncated {\bf at} the first non-trivial order, in CS90 is shown that the quality of the fit to the exact result is quite good over a wide range of energy and ellipticity. On this basis it is natural to presume that higher-order truncations would improve the quality of the prediction.
   
   However, even the simplest version of the Prendergast (\cite{prende}) method has two problematic aspects: 1) the choice of the dominant harmonic in the trigonometric series has to be made on the basis of some knowledge about the orbit type under study; 2) the determination of the coefficients in the series, which depend on the parameters of the system and on initial conditions, originates from a non-linear algebraic system the solution of which must in general be performed numerically. This second aspect spoils the approach of much of its simplicity, even more if we aim at higher order truncations and consider the growth of the number of unknown coefficients.
   
   In this paper we would like to explore the link between the Prendergast-Contopoulos approach and the approximation of orbital solutions found with a resonant normal form. The motivation for this study stems from the idea of rooting a simplified version of the rational solution method into the frame of a modified normalization algorithm in order to devise a completely analytical approach. In fact it has recently proposed (Pucacco et al. \cite{PBB}) to exploit a resummation technique based on continued fractions to speed up the convergence of series obtained in the framework of normal form perturbation theory. This technique is able to extend the quality of predictions concerning the instability of normal modes and consequent bifurcations of families of boxlets (Belmonte et al. \cite{BBP2}).
   
 In analogy with  CS90 we apply this approach to investigate periodic orbits in the logarithmic potential (Binney \& Tremaine \cite{BT}). We find analytical solutions of the equations of motion for the normal modes and the main low-order boxlets (`loops' and `bananas'). By inverting the normalizing transformation of coordinates, these solutions are either in the form of standard truncated power series or in a rational form constructed by a continued fraction truncated at the same order of the series. Knowing the `normal form' approximating the system under study, the procedure of creating those solutions is straightforward and does not require any further approximation or numerics.
 
 We show that the analytic rational solutions obtained in this way offer a degree of reliability comparable, where data are available, to those of the semi-analytic treatment based on the Prendergast-Contopoulos approach. Both loops and bananas are quite well reconstructed in shape and dimension. We extend the analysis in CS90 also to check the energy conservation along the boxlets: it turns out that, whether for normal modes energy is conserved within a few percent, for loops and bananas, at this level of approximation, it is not easy to go below 10\%.
 
 The plan of the paper is as follows: in Section 2 we briefly recall the method to construct normal forms for the logarithmic potential, relegating to the Appendix the explicit expressions of the 1:1 and 1:2 Hamiltonian and generating function. In Section 3 we analyze the approximation of the major-axis orbit and in Sections 4 and 5 we do the same respectively for the loop and banana families. In Section 6 we sketch the conclusions.
   

\section{Normal forms for the logarithmic potential}

We investigate the dynamics in the potential
\be\label{v1}
V=\frac12 \log\left(R^{2} + x^2 + \frac{y^2}{q^2}\right).\ee
For every finite values of the ``core radius'' $R$, the choice $R=1$ can be done without any loss of generality. However, it is also of relevance the singular limit $R\rightarrow0$ associated to a central density cusp (Miralda-Escud\'e \& Schwarzschild \cite{mes}). With the choice $R=1$, the energy $E$ may take any non-negative value. Otherwise, the singular limit is `scale-free' and the dynamics are the same at every energy. The parameter $q$ gives the ``ellipticity'' of the figure and ranges in the interval
 \be\label{rangeq}
 0.6 \le q \le 1. \ee
 Lower values of $q$ can in principle be considered but correspond to an unphysical density distribution. Values greater than unity are included in the treatment by reversing the role of the coordinate axes. 

Normal forms for the Hamiltonian system corresponding to the potential (\ref{v1}) are constructed with standard methods (Boccaletti \& Pucacco \cite{DB}, Giorgilli \cite{gior}) and have been used to determine the main features of its orbit structure (Belmonte et al. \cite{BBP1,BBP2}). The starting point is the series expansion of (\ref{v1}) around the origin
\be\label{ELP}
 V = \frac12 s -  \frac{1 }{2 \cdot 2}  s^{2} + \frac{1 }{2 \cdot 3}  s^{3} - \frac{1 }{2 \cdot 4}  s^{4} + \dots
 \ee
where
\be
s = x^2 + \frac{y^2}{q^2}.\ee
Here we briefly resume the procedure just with the purpose of fixing notations. After a scaling transformation
\be
p_y \longrightarrow {\sqrt{q}} \, p_{y}, \quad  
y \longrightarrow y/\sqrt{q},\ee
the original Hamiltonian 
\be\label{Horig}
   H(p_x,p_y,x,y)= \frac{1}{2}(p_x^2+p_y^2/q) + V(s(x,y))
\ee
is subject to a canonical transformation to new variables $P_{X},P_{Y},X,Y$, such that 
\begin{equation}\label{HK}
     K(P_{X},P_{Y},X,Y)=\sum_{n=0}^{N}K_n ,
  \end{equation}
with the prescription ($K$ in `normal form')
\be\label{NFD}
\{K_0,K\}=0.
\ee
In these and subsequent formulas we adopt the convention of labeling the first term in the expansion with the index zero: in general, the `zero order' terms are quadratic homogeneous polynomials and terms of {\it order} {\it n} are polynomials of degree $n+2$. The zero order (unperturbed) Hamiltonian,  
\be\label{Hzero}
K_{0} \equiv H_{0} = \frac12 (P_X^2 + X^2) + \frac{1}{2q} (P_Y^2  +Y^2),
\ee
with `unperturbed' frequencies $\omega_1=1,\omega_2 = 1/q$, plays, through the fundamental equation (\ref{NFD}), the double role of determining the specific form of the transformation and of assuming the status of second integral of motion. 

The generating function of the transformation is a series of the form 
\be\label{gene}
G=G_{1}+G_{2}+...\ee 
and, since the procedure is based on working at each order with quantities determined at lower orders, the normalization algorithm proceeds by steps up to the `truncation' order $N$. At each step $n$ (with $ 1 \le n \le N$), the series are `upgraded' expressing them in the new variables found with the normalizing transformation. In the Appendix A we detail the expression of the normal forms and the generating function we will need in the following.

It is customary to refer to the series constructed in this way as {\it Birkhoff} normal forms. The presence of terms with small denominators in the expansion, forbids in general their convergence. It is therefore more effective to work since the starting point with {\it resonant normal forms} (Sanders, Verhulst \& Murdock \cite{SV}, Gustavson \cite{gu}), which are still non-convergent, but have the advantage of avoiding the small divisors associated to a particular resonance. To catch the main features of the orbital structure, we therefore approximate the frequencies with a rational number plus a small `detuning' (Contopoulos \& Moutsoulas \cite{CM}, de Zeeuw \& Merritt \cite{zm})
\be\label{DET}
\frac{\omega_1}{\omega_2} = \frac{m_1}{m_2} + \delta. \ee
We speak of a {\it detuned} ($m_1$:$m_2$) {\it resonance}, with $m_1+m_2$ the {\it order} of the resonance. Each resonance allows us to describe a set of possible periodic orbits appearing in the dynamics: we have the 1:1 `loop', the 1:2 `banana', the 2:3 `fish' and so forth (Miralda-Escud\'e \& Schwarzschild \cite{mes}). Each of them, if stable, is surrounded by a family of quasi-periodic orbits usually inheriting the same nickname. 

A very conservative strategy can be that of truncating at the lowest order $N_{\rm min}$ adequate to convey some non-trivial information on the system. In the resonant case, it can be shown (Tuwankotta \& Verhulst \cite{tv}) that the lowest order to be included in the normal form in order to capture the main effects of the $m_1$:$m_2$ resonance with double reflection symmetries is 
\be\label{Nmin}
N_{\rm min} = 2\times(m_1+m_2-1).\ee  
Using `action-angle'--{\it like} variables $\vec{J}, \vec{\theta}$ defined through the transformation
\ba
X &=& \sqrt{2 J_1} \cos \theta_1,\quad
Y  = \sqrt{2 J_2} \cos \theta_2,\label{AAV1}\\
P_X &=& \sqrt{2 J_1} \sin \theta_1,\quad
P_Y = \sqrt{2 J_2} \sin \theta_2,\label{AAV2}\ea 
the typical structure of the doubly-symmetric resonant normal form truncated at $N_{\rm min}$ is (Sanders, Verhulst \& Murdock \cite{SV}, Contopoulos \cite{c1})
\begin{eqnarray}\label{GNF}
K&=&m_{1} J_1+m_{2} J_2+ \sum_{k=2}^{m_1+m_2} {\cal P}^{(k)}(J_1,J_2)+\nonumber \\
&&a_{m_1 m_2} J_1^{m_{2}} J_2^{m_{1}} \cos [2(m_{2} \theta_{1}- m_{1} \theta_{2})], 
\end{eqnarray}
where ${\cal P}^{(k)}$ are homogeneous polynomials of degree $k$ whose coefficients may depend on $\delta$ and the constant $a_{m_1 m_2}(q)$ is the only marker of the resonance. In these variables the second integral is 
\be\label{cale}
{\cal E}=m_{1} J_1+m_{2} J_2\ee
and the angles appear only through the resonant combination
\be\label{psi}
\psi=m_{2} \theta_{1}- m_{1} \theta_{2}.\ee
For a given resonance, the last two statements remain true for arbitrary $N>N_{\rm min}$. Introducing the variable conjugate to $\psi$,
\be\label{calr}
{\cal R}=m_{2} J_1-m_{1} J_2,\ee
the new Hamiltonian can be expressed in the {\it reduced} form $K({\cal R}, \psi; {\cal E},q)$, that is a family of 1-dof systems parametrized by ${\cal E}$ and $\delta$. 

We are interested in the solution of the equations of motion. For a non-resonant (Birkhoff) normal form the problem is easily solved: the coefficient $a_{m_1 m_2}$ vanishes, so $K$ lacks the term containing angles. Therefore the $\vec{J}$ are `true' conserved actions and the solutions are
\be\label{SF}
X (t) = \sqrt{2 J_1} \cos \kappa_1 t ,\quad
Y (t) = \sqrt{2 J_2} \cos (\kappa_2 t + \theta_0),\ee
where
\be
\vec{\kappa} = \nabla_{\vec{J}} K\ee
is the frequency vector and $\theta_0$ a suitable phase shift. 

In the resonant case instead, it is no more possible to write the solutions in closed form. It is true that the dynamics described by the 1-dof Hamiltonian $K({\cal R}, \psi; {\cal E},q)$ are always integrable, but, in general, the solutions cannot be written in terms of elementary functions. However, solutions can still be worked out in the case of the {\it main periodic orbits}, for which $\vec{J}, \vec{\theta}$ are again true action-angle variables. There are two kinds of periodic orbits that can be easily identified:
\begin{enumerate}
      \item The {\it normal modes} for which one of the $\vec{J}$ vanishes.
         
      \item  The {\it periodic orbits in general position} characterized by a {\it fixed} relation between the two angles, $m_{2} \theta_{1}- m_{1} \theta_{2} \equiv \theta_0$.
       \end{enumerate}
In both cases, it is easy to check that the solutions retain a form analogous to (\ref{SF}) with known expressions of the actions and the frequencies in terms of ${\cal E}, q$ and such that $\kappa_1 / \kappa_2 = m_1 / m_2$.

By using the generating function (\ref{gene}), the solutions in terms of standard `physical' coordinates can be recovered (a part for possible scaling factors) inverting the canonical transformation defined by (\ref{TNFD}) and (\ref{eqn:OperD-F}). As discussed in the Appendix, the expansion (\ref{gene}) is actually composed of even-order terms only. Since in our applications we will treat the 1:1 and 1:2 symmetric resonances, we have from (\ref{Nmin}) that at most $N_{\rm min} = 4$ so that the transformation back to the physical coordinates expressed as a series of the form
\be
\vec{x} (t) = \vec{x}_{1} + \vec{x}_{2} + \vec{x}_{3} +...\ee   
is explicitly given by
\ba
\vec{x}_{1} &=& \vec{X}, \label{x1}\\
\vec{x}_{2} &=& 0, \\
\vec{x}_{3} &=& L_{2}(\vec{X})=\{G_{2},\vec{X}\}, \label{x3} \\
\vec{x}_{4} &=& 0, \\
\vec{x}_{5} &=& L_{4}(\vec{X}) + {\scriptstyle\frac12} L^{2}_{2}(\vec{X})=
\{G_{4},\vec{X}\} + {\scriptstyle\frac12} \{G_{2},\{G_{2},\vec{X}\}\}. \label{x5} \ea
We again remark that the vanishing of terms of even degree is related to the double reflection symmetry embodied in the normal form. From a knowledge of the normalized solutions (\ref{SF}), we can therefore construct power series approximate solutions of the equations of motion of the original system
\be
\frac{d^{2} \vec{x}}{dt^{2}} = -\nabla_{\vec{x}} V.\ee

We are investigating a non-integrable system. This implies that any perturbation approach to cope with its dynamics is deemed to fail, since it produces series which do not converge in general. On the other hand, what the normal form provides us is an efficient way to construct series with an {\it asymptotic} character: this means that at some point we should reach an `optimal' value for the expansion order $N_{\rm opt}$ (hopefully $>N_{\rm min}$) which gives the best possible result (Efthymiopoulos et al. \cite{ECG}). The optimal order depends on the size of the phase-space region we are interested in. The bigger this region is, the lower are the value of $N_{\rm opt}$ and the accuracy of the approximation. In galactic dynamics (contrary to what happens in celestial mechanics) it is usually preferable to get an overall picture of the dynamics giving up extreme accuracy, so that truncating at $N_{\rm min}$ seems a reasonable choice. To verify if this conjecture is actually tenable is another aim of the present work.

\section{Axial orbits}

In systems of the form (\ref{Horig}) the orbits along the symmetry axes are simple periodic orbits. It can be readily verified that these orbits correspond to the two normal modes for which either $J_1$ or $J_2$ vanish. If the axial orbit is stable it parents a family of `box' orbits. A case that is both representative of the state of affairs and useful in galactic applications is that of the stability of the  {\it x}-axis periodic orbit (the `major-axis orbit', if $q$ is in the range (\ref{rangeq})). Among possible bifurcations from it, the most prominent is usually that due to the 1:2 resonance between the frequency of oscillation along the orbit and that of a normal perturbation, producing the `banana' and `anti-banana' orbits (Miralda-Escud\'e \& Schwarzschild \cite{mes}). Therefore, to get explicit solutions both for the major-axis orbit and the stable bananas (the `pendulum-like family' in the denomination of CS90, see Section 5 below) we use the 1:2 symmetric normal form. 

From the expression of $K$ reported in the Appendix, we get on the normal mode $J_2 = 0$,
\be\label{K4Ia}
K_{A} =2q J_1 - \frac34 q J_1^2 + \frac12 q \left(\frac53 - \frac{17}{4} q(q-1) \right) J_1^3.\ee
The value of the action can be computed by using the rescaling (\ref{newE}), namely $K_{A} =2qE$ and inverting the series. $E$ is the original `physical' energy and can also be expressed by means of the amplitude $A$ of the axial orbit
\be\label{Sconvx}
E = \frac12 \log (1+A^{2}).\ee 
The frequency is given by the usual differentiation 
\be\kappa_{1} = \frac1{2q}\frac{\partial K_{A}}{\partial J_{1}},\ee
where the rescaling of the energy is taken into account in order to be able to use $t$ as the physical time. Therefore, in the normalization variables we have a solution of the form (\ref{SF}) with  $Y=0$ and (Belmonte et al. \cite{BBP2})
\ba\label{SA}
J_{1} &=& E + \frac38 E^2 + \frac{25}{192} E^3, \label{SA1}\\
\kappa_{1} &=& 1 - \frac{3}{4} E +
                          \frac{11}{64} E^{2}. \label{SA2}
\ea
Inserting this solution in the transformation formulas (\ref{x1}--\ref{x5}) and exploiting the terms of the generating function of (\ref{g12-2}) and  (\ref{g12-4}), after a little of computer algebra we get
\ba
x_{1} &=& A \cos \kappa_1 t , \label{xm1}\\
x_{3} &=& \frac{A^{3}}{32} \left( \cos \kappa_1 t - \cos 3 \kappa_1 t \right), \label{xm3}\\
x_{5} &=& \frac{A^{5}}{64} \left( - \frac{59}{48} \cos \kappa_1 t + \cos 3 \kappa_1 t +
\frac{11}{48} \cos 5 \kappa_1 t\right), \label{xm5}\ea
This result coincides with that obtained by Scuflaire (\cite{scu}) with an independent approach based on the Poincar\'e--Lindstedt method and provides the explicit time evolution of an oscillation starting at rest from $x(0) = A$. 
   
   \begin{table}
      \caption[]{Relative energy variations along the major-axis orbit with different analytic predictions.}
         \label{T1}
     $$ 
         \begin{array}{c||c||c|c||c|c}
            \hline
            \noalign{\smallskip}
            E      &  A & x^{(3)}_{NF} & x^{(5)}_{NF} & x^{(3)}_{CF} & x^{(5)}_{CF}\\
            \noalign{\smallskip}
            \hline
            \noalign{\smallskip}
            0.1         & 0.47 & 0.0013   & 0.0005 & 0.002  & 0.0001\\
            0.2         & 0.70 & 0.005     & 0.005   & 0.009  & 0.001\\
            0.3         & 0.91 & 0.011     & 0.020   & 0.022  & 0.003\\
            0.4         & 1.11 & 0.017     & 0.058   & 0.046  & 0.008\\
            0.5         & 1.31 & 0.02       & 0.13     & 0.08    & 0.02        \\
            0.6         & 1.52 & 0.03       & 0.25     & 0.16    & 0.03        \\
            0.7         & 1.75 & 0.08       & 0.42     & 0.29    & 0.05        \\
            0.8         & 1.99 & 0.15       & 0.70     & 0.55    & 0.07        \\
            0.9         & 2.25 & 0.25       & 0.95     & --         & 0.09        \\
            1.0         & 2.53 & 0.35       & --     & --              & 0.12        \\
            \noalign{\smallskip}
            \hline
         \end{array}
     $$ 
   \end{table}

To evaluate the quality of the approximation, a simple method is to follow the energy variation along the solution in the true potential (\ref{v1}). We therefore compute
\be
{\tilde E}(t) = \frac12 \left(\frac{dx}{dt}\right)^{2} + \frac12 \log (1+x(t)^{2})\ee
and compare it with the given value of $E$ fixed by (\ref{Sconvx}) for various amplitudes. To understand the question of the optimal order we can choose two different truncations of the prediction obtained with the normal form:
\ba
x^{(3)}_{NF} &=& x_{1}+x_{3}, \label{NF3}\\
x^{(5)}_{NF} &=& x_{1}+x_{3}+x_{5} \label{NF5}\ea
and compute the quantity
\be
\frac{\Delta E}{E} = \frac{{\tilde E (t)} - E}{E}.\ee

   \begin{figure}
   \centering
\includegraphics[width=9cm]{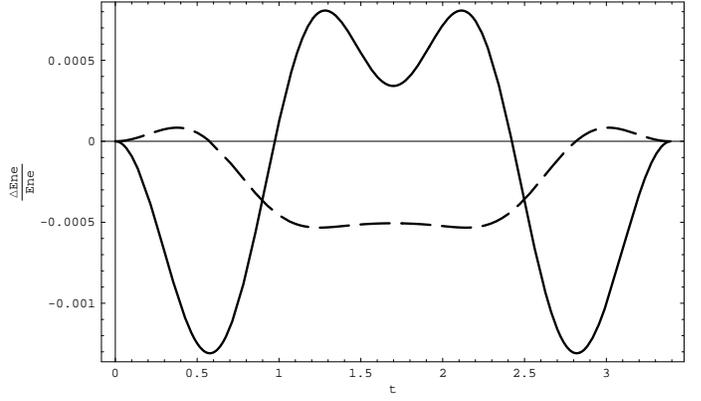}
      \caption{Relative energy error along the major-axis orbit with two different truncations of the normal form at 
           $E=0.1$.   }
         \label{nf1}
   \end{figure}
  \begin{figure}
   \centering
\includegraphics[width=9cm]{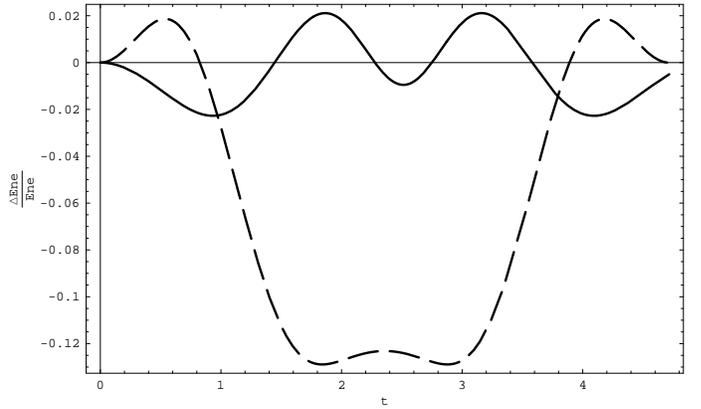}
      \caption{Relative energy error along the major-axis orbit with two different truncations of the normal form at 
           $E=0.5$.   }
         \label{nf2}
   \end{figure}
In Fig. \ref{nf1} we plot $\Delta E/E$ for $E=0.1$ (corresponding to an  amplitude $A=0.47$) over a half period: the curves repeat themselves in the subsequent half period. The solid line is computed with $x^{(3)}_{NF}$ and the dashed line with $x^{(5)}_{NF}$: the relative error in the energy conservation is almost three times smaller with the higher truncation and as low as $0.05 \%$. However, from Fig. \ref{nf2} we see that with $E=0.5$ ($A=1.31$) the situation is upset: the lower order truncation (which corresponds just to the first non-zero term in the normal form) gives an error at least five times smaller than that with the higher truncation. We deduce that somewhere between the two energy levels the optimal order decreases by two and verify that, to get informations about an orbit 3 times bigger, we must accept a relative error of a few percent. In fact, in Table (\ref{T1}), we list the maximum absolute energy variation over a half period for various values of $E$ and see that the optimal order is $\ge 4$ up to $E=0.2$: for greater values of the energy the optimal order is just 2, namely with $x^{(3)}_{NF}$ we have the best result.

Once reached the optimal order, it can be disappointing to discard terms coming from a costly high-order computation. There are however other rules for `summing' divergent series which make use of all terms (Bender \& Orszag, \cite{BO}), like the construction of Pad\`e approximant. A related approach is that of constructing {\it continued fractions}: successive approximants obtained by truncating the fraction at various order may give an improvement in the asymptotic convergence with respect to the original series (Khovanskii, \cite{KHO}). From the normal form series (\ref{NF3},\ref{NF5}) we may compute the truncated fractions
\ba
x^{(3)}_{CF} &=& \frac{x_{1}}{1-x_{3}/x_{1}}, \label{CF3}\\
x^{(5)}_{CF} &=& \frac{x_{1}}{1-\frac{x_{3}/x_{1}}{1+\frac{x_{3}^{2} - x_{1}x_{5}}{x_{1}x_{3}}}}. \label{CF5}\ea
These approximations produce rational solutions and is therefore natural to think to a relation with the Prendergast-Contopoulos approach of CS90. By using the explicit forms of (\ref{xm1}--\ref{xm5}) we get
\ba
x^{(3)}_{CF} &=& \frac{ A \cos \kappa_1 t  }{1+\frac{A^{2}}{16} (1 - \cos 2 \kappa_1 t) }, \label{CF3E}\\
x^{(5)}_{CF} &=& A \cos \kappa_1 t \
\frac{1+A^{2} (\frac{65}{96} +  \frac{1}{6} \cos 2 \kappa_1 t) }
       {1+A^{2} (\frac{59}{96} +  \frac{11}{48} \cos 2 \kappa_1 t) }
\ea
and it turns out that the expression of $x^{(3)}_{CF}$ has the same structure of the trial rational approximation used in CS90:
\be\label{PC1}
\frac{ {\tilde A} \cos \kappa_1 t  }{1+ B \cos 2 \kappa_1 t }.\ee
The main difference is that in $x^{(3)}_{CF}$ the parameters $A$ and $\kappa_1$ are known in analytic form by 
(\ref{Sconvx}) and (\ref{SA2}), whereas in (\ref{PC1}), ${\tilde A}, B $ and $\kappa_1$ have to be {\it numerically} computed by solving a nonlinear algebraic system obtained by inserting the trial solution into the equations of motion.

  \begin{figure}
   \centering
\includegraphics[width=9cm]{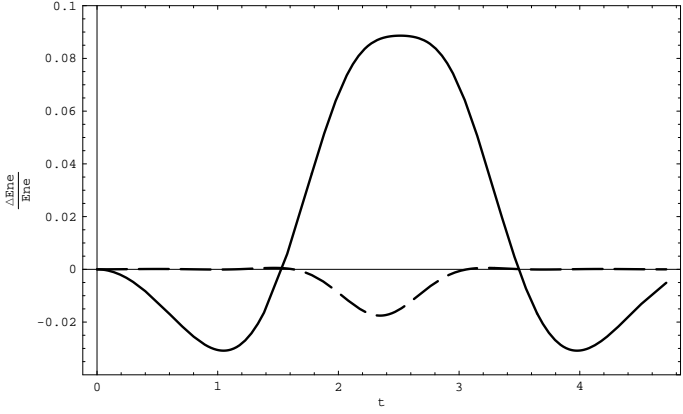}
      \caption{Relative energy error along the major-axis orbit with two different truncations of the continued fraction at 
           $E=0.5$.   }
         \label{nf3}
   \end{figure}
      
 In  Fig. \ref{nf3} we plot the same quantities of Fig. \ref{nf2} now obtained via the continued fraction truncations: the solid line is computed with $x^{(3)}_{CF}$ and the dashed line with $x^{(5)}_{CF}$. At this energy level, the prediction with $x^{(5)}_{CF}$ starts to outperform $x^{(3)}_{NF}$. From Table \ref{T1} we see that the performance of $x^{(5)}_{CF}$ is the best when going to higher energies and is at least as good as that of CS90 in the same range of energy.
 
 \section{Loop orbits} 
 
As a first example of boxlet we treat the `loop' orbits for which, to get explicit solutions, we can use the 1:1 symmetric normal form. For moderate ellipticities $(q>0.7)$, loops ensue as the lowest energy bifurcation due to the 1:1 resonance between the frequency of oscillation along the short ($y$-axis) periodic orbit and that of a normal perturbation (Miralda-Escud\'e \& Schwarzschild \cite{mes}).  From (\ref{Nmin}), for the 1:1 resonance we have  $N_{\rm min} = 2$ so that    a normal form truncated at $K_{2}$ is already able to produce loops. The bifurcation curve in the $(q,E)$-plane starts from the point $(1,0)$ (Scuflaire \cite{scu}; Belmonte et al. \cite{BBP2}) and can be expressed as the series 
\be\label{ecq11}
E_{\rm c}(q) = 2 \left(1-q \right)  + \left(1-q \right)^2 - \frac56 \left(1-q \right)^{3}...\ee 
if the normal form is truncated at progressively higher orders. 
We limit ourselves to the case $q=0.9$ with transition energy $E_{\rm c}(0.9)=0.21$ and investigate the analytic prediction of the theory by fixing the energy level at $E=1$: with suitable rescaling, these are the same values of the parameters used in CS90. 
  
In the normalization variables we have a solution of the form (\ref{SF}) with  
  \be\label{loopSF}
X (t) = \sqrt{2 J_1} \cos \kappa_{L} t ,\quad
Y (t) = \sqrt{2 J_2} \sin \kappa_{L} t ,\ee
with phase shift $\theta_0=\pi/2$. The actions and frequencies can be obtained from the following procedure: starting from the normal form (\ref{K11-0}--\ref{K11-2}), we determine the fixed points of the reduced Hamiltonian $K({\cal R}, \psi; {\cal E},q)$ with
\ba
{\cal E}&=&J_1+ J_2, \label{cale11} \\
\psi&=&\theta_{1}- \theta_{2}, \label{psi11} \\
{\cal R}&=&J_1- J_2. \label{calr11} \ea
The fixed point corresponding to the loop is given by $\psi=\pi/2$ and
\ba
J_{1(L)} ({\cal E},q) &=&\left({\cal E} + {\cal R}_{L}({\cal E},q)\right)/2, \label{J1-11}\\ 
J_{2(L)} ({\cal E},q) &=&\left({\cal E} - {\cal R}_{L}({\cal E},q)\right)/2, \label{J2-11} \ea
where ${\cal R}_{L}({\cal E},q)$ is the solution of the algebraic equation
\be\label{psid11}
\frac{\partial K}{\partial \cal R} \left({\cal R},\frac{\pi}{2}; {\cal E},q\right)=0.\ee
The frequency is then given by
\be\label{kl}
\kappa_{L} = \frac1{q} \frac{\partial K}{\partial J_{1(L)}} \left({\cal R}_{L},\frac{\pi}{2}; {\cal E},q\right).\ee
Explicit expressions of actions and frequencies are (Belmonte et al. \cite{BBP2})
\ba
J_{1(L)} ({\cal E},q) &=&\frac{(3-q) {\cal E} + 4 q (q-1)}{3 q^{2}-2q+3}, \label{LOOP1}\\ 
J_{2(L)} ({\cal E},q) &=&\frac{q(3q-1) {\cal E} - 4 q (q-1)}{3 q^{2}-2q+3}, \label{LOOP2} \\
\kappa_{L} &=& \frac1{q} \left( 1 - \frac{3}{4 q} {\cal E} \right). \label{LOOP3} 
\ea

   \begin{figure}
   \centering
\includegraphics[width=9cm]{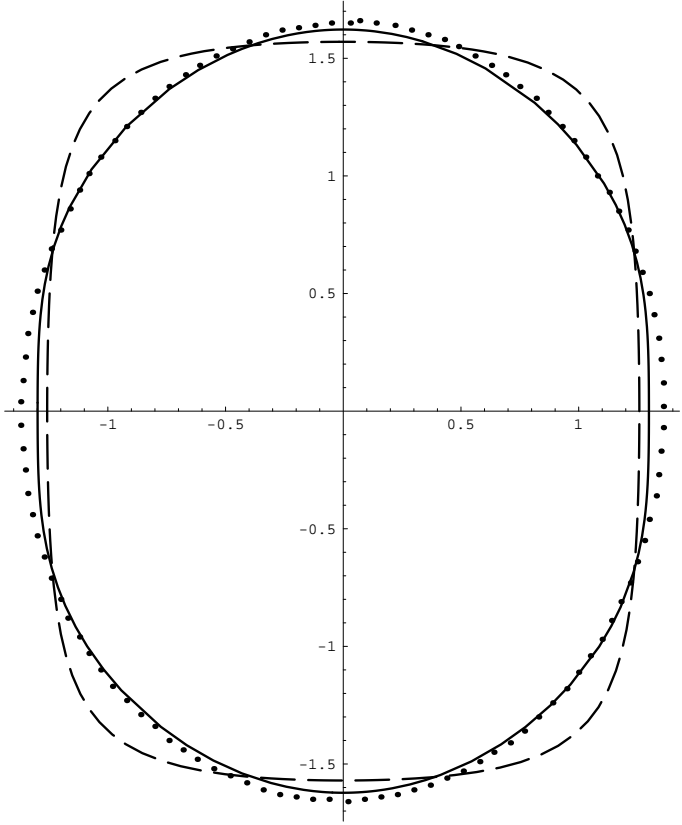}
      \caption{An orbit of the loop family at
           $E=1.0$ for $q=0.9$: dots correspond to the numerical solution; the continuous line corresponds to the prediction given by the continued fraction truncated at order 3; the dashed line that  given by the normal form truncated at order 3  }
         \label{l3}
   \end{figure}
   
    \begin{figure}
   \centering
\includegraphics[width=9cm]{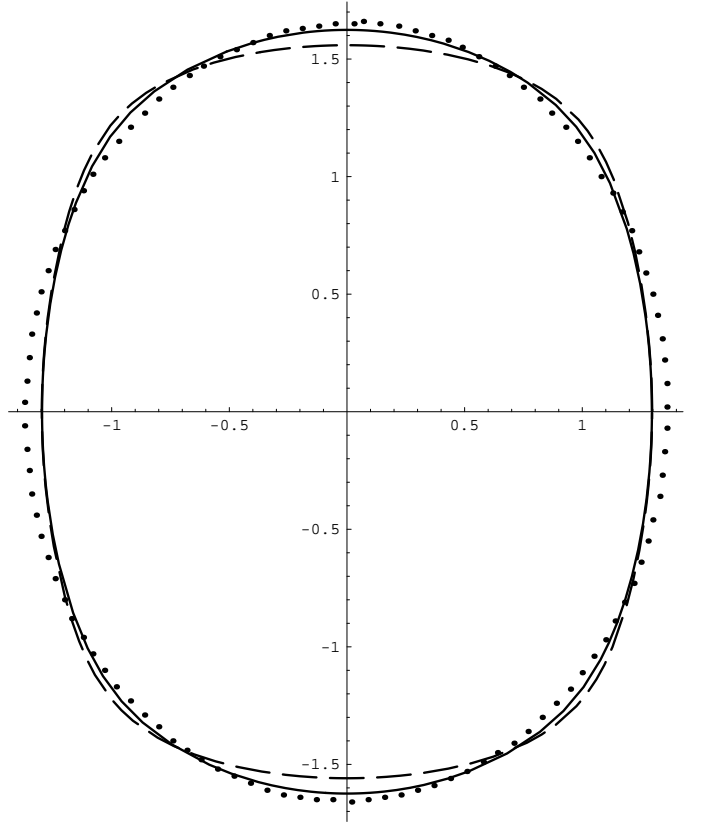}
      \caption{The same orbit of the previous figure (dots) compared with the predictions truncated at order 5 (continued fraction, continuous line; normal form, dashed line).   }
         \label{l5}
   \end{figure}
  
   \begin{figure}
   \centering
\includegraphics[width=9cm]{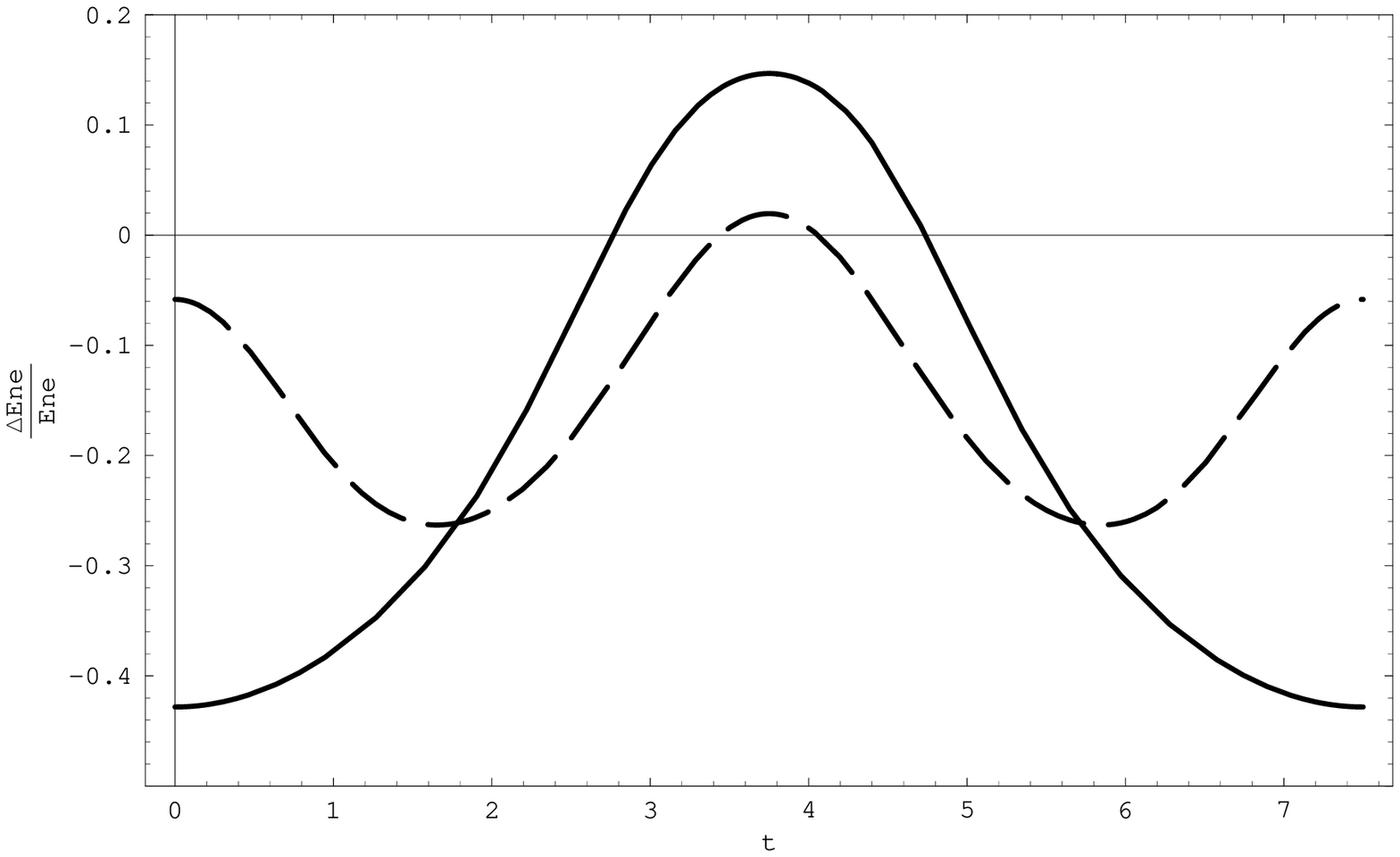}
      \caption{Relative energy error along the loop orbit with two different truncations of the continued fraction. }
         \label{lene}
   \end{figure}

For the solutions in the physical variables, we first work out the transformations (\ref{x1}--\ref{x3}) with the generating function (\ref{g11-2}) obtaining
\ba
x_{1} &=& \sqrt{2 J_{1(L)}} \cos \kappa_L t , \label{xl1}\\
x_{3} &=& \frac{1}{16} \sqrt{2 J_{1(L)}} \cos \kappa_L t \times \nonumber \\
&& \left(7(J_{2(L)} + q J_{1(L)}) -2(2 J_{2(L)} + q J_{1(L)}) \cos 2 \kappa_L t \right), \label{xl3} \\
y_{1} &=& \sqrt{2 J_{2(L)}} \sin \kappa_L t , \label{yl1}\\
y_{3} &=& \frac{1}{16 q} \sqrt{2 J_{2(L)}} \sin \kappa_L t  \times \nonumber \\
&& \left(7(J_{2(L)} + q J_{1(L)}) + 2 ( J_{2(L)} + 2q J_{1(L)} )\cos 2 \kappa_L t \right). \label{yl3}\ea
We observe that, already in the first higher-order terms both actions appear, to testify the strong coupling between the degrees of freedom. 
The inversion of the series 
\be
E=\frac1{q} K_{L}({\cal E},q)=\frac1{q} K\left({\cal R}_{L}({\cal E},q),\frac{\pi}{2}; {\cal E},q\right)\ee
allows us to express actions and frequencies in terms of the physical energy. However, using the exact solutions (\ref{LOOP1}--\ref{LOOP2}) to evaluate (\ref{xl1}--\ref{yl3}) would result in messy expressions hindering the procedure of resummation with the continuous fraction. Therefore, in analogy with the series written for the axial orbit, a separation of terms of given orders is necessary and is obtained by linearly expanding the actions in the form
\ba
J_{1(L)}&=&a(E-E_{\rm c}),\label{j1loop}\\
J_{2(L)}&=&b+c(E-E_{\rm c}).\label{j2loop}\ea
The first expansion provides an approximate value of $J_{1}$ above the bifurcation energy and clearly must be considered zero for $E<E_{\rm c}$. The second one joins in $E=E_{\rm c}$ with the corresponding expression on the normal mode. 
Inserting these into the solutions above and expanding in powers of $E - E_{\rm c}$ we are able to group terms according to their order. Clearly, this grouping does not affect the series themselves (namely $x^{(3)}_{NF}$ and $y^{(3)}_{NF}$) but rather it influences the computation of the truncated fractions: we get
\ba
x^{(3)}_{CF} &=& 5 \sqrt{2a(E-E_{\rm c})} \ \cos \kappa_ L t \ \frac{(16 + 7 b - 4 b \cos 2 \kappa_L t)^{2}}{A_{1} + A_{2} \cos 2 \kappa_L t }, \label{XCF3L}\\
y^{(3)}_{CF} &=& b^{3/2} \sin \kappa_L t \
\frac{(72 + 35 b + 10 b \cos 2 \kappa_L t)^{2}}{B_{1} + B_{2} \cos 2 \kappa_L t }, \label{YCF3L}
\ea
where
\ba
A_{1}  &=& 4 \left(160 + 70 b -(63a+70c)(E-E_{\rm c})\right),\\
A_{2}  &=& -8 \left(20 b -(9a+20c)(E-E_{\rm c})\right),\\
B_{1}  &=& 18\sqrt{2} \left(2 b (72+35b)-3(21ab+35cb+34c)(E-E_{\rm c})\right),\\
B_{2}  &=& 36 \sqrt{2} b \left(10 b -3(6a+5c)(E-E_{\rm c})\right).
\ea
For moderate values of the bifurcation energy, corresponding to large values of $q$ in the range (\ref{rangeq}), a simple approximation is given by the linear term in (\ref{ecq11}), $E_{\rm c}=2(1-q)$. At this level of approximation the constants appearing in the above solutions are
\ba
a&=&\frac32  -q ,\\
b&=&2\left(1-q  \right) = E_{\rm c}, \\
c&=&\frac{q}2\ea
and can be used to plot the orbits and compare them with numerical computations. 
With the choice of the parameters mentioned above, in Fig. \ref{l3} we compare a numerical computation of the loop orbit (dots) with the analytic predictions given by $\vec{x}^{(3)}_{NF}$ (dashed line) and $\vec{x}^{(3)}_{CF}$ (continuous line). It appears clear how the rational solution coming from the continued fraction truncated at order 3 is overall quite accurate in locating both the shape and the extrema of the orbit and overtakes the prediction with the standard truncated series. In Fig. \ref{l5} we compare the numerical computation of the same orbit (dots) with the analytic predictions given by $\vec{x}^{(5)}_{NF}$ (dashed line) and $\vec{x}^{(5)}_{CF}$ (continuous line). The prediction with the higher order truncation does not appear so much better as far as the shape of the loop. However, in Fig. \ref{lene} we may compare the plots of $\Delta E/E$ versus time for the two truncations (order 3, continuous line; order 5, dashed line) over a half period: the relative error is some 25\% at order 5 contrary to more than  40\% with $\vec{x}^{(3)}_{CF}$.

\section{Pendulum-like (banana) orbits}

  The bifurcation of the banana orbit from the major-axis occurs along a curve in the $(q,E)$-plane starting from the point $(1/2,0)$ (Scuflaire \cite{scu}; Belmonte et al. \cite{BBP2}). It can be expressed as the series 
\be\label{ecq12}
E_{\rm c}(q) = 8 \left(q - \scriptstyle\frac12 \right)  - \frac{20} 3 \left(q - \scriptstyle\frac12 \right)^2 + \frac{268}{9} \left(q - \scriptstyle\frac12 \right)^{3}...\ee 
As before we adopt the same values of the parameters (up to a suitable rescaling) used in CS90: the ellipticity is $q=0.6$ with transition energy $E_{\rm c}(0.6)=0.76$ and the energy level is fixed at $E=1.15$.
  
  In the normalization variables we have a solution of the form (\ref{SF}) with  
  \be\label{banSF}
X (t) = \sqrt{2 J_1} \cos \kappa_{B} t ,\quad
Y (t) = \sqrt{2 J_2} \cos 2 \kappa_{B} t .\ee
To find actions and frequencies, starting from the normal form (\ref{K12-0}--\ref{K12-4}), we determine the fixed points of the reduced Hamiltonian $K({\cal R}, \psi; {\cal E},q)$ with
\ba
{\cal E}&=&J_1+ 2 J_2, \label{cale12} \\
\psi&=&2 \theta_{1}- \theta_{2}, \label{psi12} \\
{\cal R}&=&2 J_1- J_2. \label{calr12} \ea
The fixed point corresponding to the banana is given by $\psi=0$ and
\ba
J_{1(B)} ({\cal E},q) &=&\left({\cal E} + 2 {\cal R}_{B}({\cal E},q)\right)/5, \label{J1-12}\\ 
J_{2(B)} ({\cal E},q) &=&\left(2{\cal E} - {\cal R}_{B}({\cal E},q)\right)/5, \label{J2-12} \ea
where ${\cal R}_{B}({\cal E},q)$ is the solution of the algebraic equation
\be\label{psidot}
\frac{\partial K}{\partial \cal R} ({\cal R},0; {\cal E},q)=0.\ee
The frequency is then given by
\be\label{kb}
\kappa_{B} = \frac1{2q} \frac{\partial K}{\partial J_{1(B)}} ({\cal R}_{B},0; {\cal E},q).\ee
${\cal R}_{B}({\cal E},q)$ and, as a consequence $J_{1(B)}, J_{2(B)}$ and $\kappa_{B}$ are quite cumbersome algebraic expressions involving ${\cal E}$ and $q$. However, simple expressions to represent orbits in the initial physical coordinates can be obtained by replacing them with some suitable series expansions. For, we first work out the transformations (\ref{x1}--\ref{x5}) with the generating function (\ref{g12-2}--\ref{g12-4}) obtaining
\ba
x_{1} &=& \sqrt{2 J_{1(B)}} \cos \kappa_B t , \label{xb1}\\
x_{3} &=& \frac{1}{24} \sqrt{2 J_{1(B)}} \cos \kappa_B t 
\big(7(2 J_{2(B)} + 3 q J_{1(B)}) \nonumber\\
&& -(4 J_{2(B)} + 6 q J_{1(B)}) \cos 2 \kappa_B t + 2 J_{2(B)} \cos 4 \kappa_B t \big), \label{xb3} \\
y_{1} &=& \sqrt{2 J_{2(B)}} \sin 2 \kappa_B t , \label{yb1}\\
y_{3} &=& \frac{1}{48 q} \sqrt{2 J_{2(B)}} \big(24 q J_{1(B)} + 
6(3 J_{2(B)} + 2 q J_{1(B)}) \cos 2 \kappa_B t \nonumber\\
&&  - 8 q J_{1(B)} \cos 4 \kappa_B t - 3 J_{2(B)} \cos 6 \kappa_B t \big) \label{yb3}\ea
and analogous expressions for $x_{5}$ and $y_{5}$. In analogy with the procedure followed for the loop orbit, a separation of terms of different low orders is useful and is obtained by linearly expanding the actions in the form
\ba
J_{1(B)}&=&a_{0}+a_{1}(E-E_{\rm c}),\\
J_{2(B)}&=&b_{1}(E-E_{\rm c}).\ea
Inserting these into the solutions above and expanding in powers of $E - E_{\rm c}$ we are able to group terms according to their order. Here again, this grouping does not affect the series themselves (namely $x^{(k)}_{NF}$ and $y^{(k)}_{NF}$, with $k=3,5$) but rather it influences the computation of the truncated fractions, namely the expressions (\ref{CF3}--\ref{CF5}) and analogous for the $y$ coordinate.

For moderate values of the bifurcation energy (and of orbital energy), corresponding to small values of $q$ in the range (\ref{rangeq}), a simple approximation is given by the linear term in (\ref{ecq12}), $E_{\rm c}=8(q-1/2)$, so that
\ba
J_{1(B)}&=&\frac12 E + \left(q - \scriptstyle\frac12 \right) \left(4 + \frac{35}{12} E \right),\label{j1b}\\
J_{2(B)}&=&\frac14 E - \left(q - \scriptstyle\frac12 \right) \left(2 - \frac{37}{24} E \right).\label{j2b}\ea
In this way, the expansions can be written as series of the form
\ba
x^{(3)}_{NF} &=& 
\sum_{j=1}^{4} A_{1j} \cos(2j-1)\kappa_B t + \nonumber\\
&&(E-E_{\rm c})\cos\kappa_B t
\sum_{j=0}^{4} A_{3j} \cos 2j \ \kappa_B t, \label{x3b}\\
y^{(3)}_{NF} &=& 
\sum_{j=0}^{5} A_{2j} \cos 2j \ \kappa_B t + \nonumber\\
&&(E-E_{\rm c})
\sum_{j=0}^{5} A_{4j} \cos 2j \ \kappa_B t, \label{y3b}
\ea
  so that, using (\ref{CF3}), one can also construct $\vec{x}^{(3)}_{CF}$. Analogously, we may proceed with the higher-order truncations $\vec{x}^{(5)}_{NF}$ from which to obtain $\vec{x}^{(5)}_{CF}$. A comparison of the structure of these predictions with the rational solutions based on the Prendergast-Contopoulos approach shows that they have the same parity in the trigonometric parts: although in the expansions (\ref{x3b}--\ref{y3b}) many more harmonics appear, this is clearly not necessarily an indication of greater accuracy.
  
\begin{figure}
   \centering
\includegraphics[width=9cm]{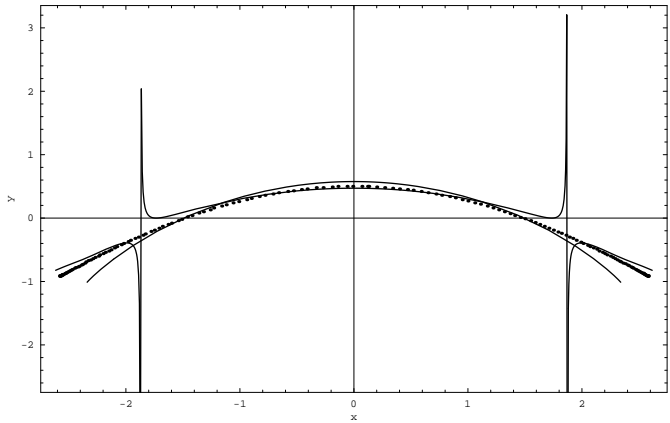}
      \caption{An orbit of the pendulum-like (banana) family at
           $E=1.15$ for $q=0.6$: the dots correspond to the numerical solution; the continuous lines correspond to the predictions truncated at order 3.   }
         \label{b3}
   \end{figure}

\begin{figure}
   \centering
\includegraphics[width=9cm]{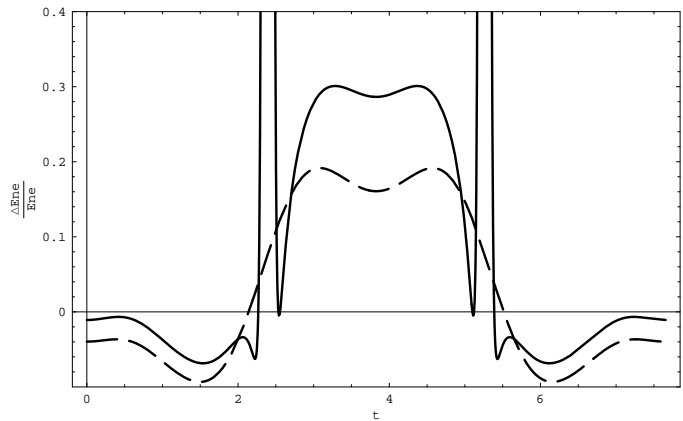}
      \caption{Relative energy error along the banana orbit with two different truncations of the continued fraction. }
         \label{bene}
   \end{figure}

With the choice of the parameters mentioned above, in Fig. \ref{b3} we compare a numerical computation of the banana orbit (dots) with the analytic predictions (continuous lines) given by $\vec{x}^{(3)}_{NF}$ and $\vec{x}^{(3)}_{CF}$. This one, the rational solution coming from the continued fraction truncated at order 3, is characterized by a pair of singularities in $y^{(3)}_{CF} (t)$ due to the presence of poles. However, the prediction is overall quite accurate in locating both the shape and the extrema of the orbit and overtakes the prediction with the standard truncated series. In Fig. \ref{bene} we plot the corresponding $\Delta E/E$ (continuous line) over a half period: the abrupt increase of the relative error is evident at the poles of the solution. 

\begin{figure}
   \centering
\includegraphics[width=9.5cm]{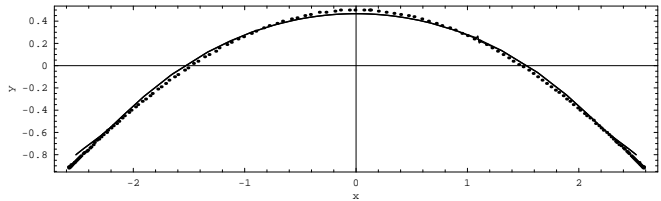}
      \caption{The same orbit of the previous figure (dots) compared with the predictions truncated at order 5 (continuous lines): for more clarity, the $y$-scale is expanded.   }
         \label{b5}
   \end{figure}

In Fig. \ref{b5} we compare the numerical computation of the same orbit (dots) with the analytic predictions given by $\vec{x}^{(5)}_{NF}$ and $\vec{x}^{(5)}_{CF}$. The two predictions now almost overlap but it could be seen a better performance of the continued fraction truncation at the extrema of the orbit. In Fig. \ref{bene} we plot the corresponding $\Delta E/E$ (dashed line) over a half period: the relative error is now less than 20\% contrary to the 30\% with $\vec{x}^{(3)}_{CF}$.

A comparison with the results of CS90 is possible only for what concerns the reconstruction of the shape and location of the orbit (we have used the same values of the parameters, when properly rescaled) and we can deduce an accuracy of our analytic predictions at least as good as that in CS90. There is no information in CS90 about the ability of their solution in conserving energy.

\section{Conclusions}

We have seen how to construct approximate solutions for the main periodic orbits in the cored logarithmic potential. The guiding line has been that of exploiting normal form expansions truncated to the first order incorporating the resonance corresponding to the given family of periodic orbits. In this way, analytic approximate solutions can be worked out with a complete algorithmic procedure. Although the attempts have always been made by truncating series to the first non-trivial orders, the solutions are definitely simple only in the case of the axial orbits (normal modes). For the low-order boxlets (loops and bananas), even the truncations at the first non-trivial order are quite cumbersome and would require the use of an algebraic manipulator. However, further simplifications can be attained if the algebraic solutions giving actions and frequencies are expanded around energy and ellipticity corresponding to the bifurcation of the family. In this case, quite simple compact expressions of the expansions can be worked out, both as standard series and as continued fractions. 

A comparison with the rational (Prendergast-Contopoulos) approach, Contopoulos (\cite{c1}), allows us to state the following conclusions: the two methods are almost equivalent for what concerns the precision of the analytic prediction when performed to the same order (Contopoulos \& Seimenis \cite{contos}). However, the normal form perturbation expansions, even if computationally heavy, are completely algorithmic and analytic at every stage, whereas the explicit evaluation of the coefficients in the rational expansions require the numerical solution of non-linear systems. We have shown that the analytic rational solutions obtained in this way offer a degree of reliability such that both loops and bananas are quite well reconstructed in shape and dimension. We have extended the analysis in Contopoulos \& Seimenis (\cite{contos}) also to check the energy conservation along the boxlets: it turns out that, whereas for normal modes energy is conserved within a few percent, for loops and bananas, at this level of approximation, it is not easy to go below 10\%. 
 
On the theoretical side, the general usefulness of rational solutions can be explained in the light of the better convergence performance of truncated continued fractions. These come into play as a resummation method of the series expansions produced in the usual way in the normalization approach. The generality of this setting allows us to envisage analogous results in the case of higher-order resonances and the corresponding higher commensurable boxlets.
 
In addition to the formal and algorithmic improvements, we remark on the relevance of this work also in relation with specific problems of galactic dynamics. The study of orbits in non-axisymmetric potentials is usually performed numerically; however, an exhaustive study with conventional integration methods is costly and difficult to interpret (Touma \& Tremaine \cite{TT}). The availability of simple and accurate analytical recipes can be quite useful in several contexts in which periodic orbits and boxlets play an important role: we mention the study of the parameter space of non-axisymmetric discs (Zhao, Carollo \& de Zeeuw \cite{ZZ}, Zhao \cite{Zhao}) and that of the orbit structure around massive black holes in galactic nuclei (Sridhar \& Touma \cite{ST}). Even more promising seems to be the possibility of getting accurate solutions for periodic orbits in the triaxial case with and without rotation, for which the analysis is still at the level of the first-order averaging method applied to the 1:1:1 resonance by de Zeeuw (\cite{dz}).

\begin{acknowledgements}
      We thank G. Contopoulos for arousing our interest in this problem.
     
\end{acknowledgements}

\begin{appendix}
\section{Resonant normal forms for the logarithmic potential}

In order to implement the normalization algorithm, the original Hamiltonian (\ref{Horig}) is rescaled according to 
\be\label{newE}{\cal H} := \frac{m_2 H}{\omega_2} = m_2 q H,\ee
so that we redefine the Hamiltonian as the series
\begin{eqnarray}\label{HS}
{\cal H} (\vec{p},\vec{x}) = \sum_{k=0}^{\infty} {\cal H}_{k} &=& \frac12 [m_1 (p_x^2+x^2) + m_2 (p_y^2+y^2)]+\nonumber \\ 
&& \frac12 m_2\delta (p_x^2+x^2) + 
\sum_{k=0}^{\infty} V_{k}(x^2,y^2) ,
\end{eqnarray}
where the detuning has been introducing as in (\ref{DET}) and the potential is the series expansion given by (\ref{ELP}). The normalization is performed with the technique of the {\it Lie transform} (Gerhard \& Saha \cite{gs}, Yanguas \cite{ya}). Considering a generating function $G$, new coordinates $\vec{P},\vec{X}$ result from the canonical transformation
  \be\label{TNFD}
  (\vec{P},\vec{X}) = M_{G} (\vec{p},\vec{x}).\ee
The {\it Lie transform operator} $M_{G}$ is defined by (Boccaletti \& Pucacco \cite{DB})
\begin{equation}\label{eqn:OperD-F}
    M_{G} \equiv \sum_{k=0}^{\infty} M_k
\end{equation}
where
\be M_0 = 1, \quad M_k = \sum_{j=1}^k \frac{j}{k} L_{G_j} M_{k-j}.\ee
The linear differential operator $L_{G}$ is defined through the Poisson bracket, $L_{G}(\cdot)=\{G,\cdot\}$ and the functions $G_j$ are the terms in the expansion of the generating function. It turns out that $G_{0}=1$ so that, in practice, the first term in its expansion can be ignored as in (\ref{gene}). The terms in the hew Hamiltonian are determined through the recursive set of linear partial differential equations 
\be\label{EHK} 
K_n= {\cal H}_n +\sum_{j=0}^{n-1}M_{n-j}{\cal H}_j ,
\;\; n=1,2, \dots
\ee 
`Solving' the equation at the $n$-th step consists of a twofold task: to find $K_{n}$ {\it and} $G_n$. We observe that, in view of the reflection symmetries of the Hamiltonian (\ref{HS}), the chain (\ref{EHK}) is composed only of members with even index and so the normal form and the generating function are composed of even-index terms only. The unperturbed part of the Hamiltonian, ${\cal H}_0$, determines the specific form of the transformation. In fact, the new Hamiltonian $K$ is said to be {\it in normal form} if, analogously to (\ref{NFD}),
\be
\{{\cal H}_0,K\}=0,
\ee
 is satisfied. In the following formulas we list the normal form and the generating function for the expansion of the logarithmic potential in the cases of the 1:1 and 1:2 resonances (Belmonte et al. \cite{BBP2}). The normal form is given in the more compact version given by using the action-angle--{\it like} variables $\vec{J}, \vec{\theta}$: the resulting expressions are in agreement with the general structure of (\ref{GNF}). For the generating function it is more useful to write the explicit version in standard $\vec{P},\vec{X}$ variables that, although cumbersome, is that exploited in the transformation back to the original $\vec{p},\vec{x}$ variables. For the 1:1 resonance the terms of the normal form are
\begin{eqnarray}
 K_{0} &=& J_1 +  J_2, \label{K11-0} \\
 K_{2} &=& \delta J_{1} - \frac{3q}{8} J_1^2 + \frac{3}{8q} J_2^2 + \frac12 J_1 J_2  + \frac14 J_1 J_2 \cos(2 \theta_1 - 2 \theta_2) , \label{K11-2} \\
 K_{4} &=& \frac{q}{4} \left(\frac53 - \frac{17q}{16} \right) J_1^3 + \frac{29}{192q^{2}} J_2^3 + \nonumber\\ 
 && \frac18 \left(\frac{39}{8} - 3q \right) J_1^2 J_2 - \frac38 \left(\frac{3}{8} - \frac{1}{q} \right) J_1 J_2^2 + \nonumber\\
 && \frac18 \left[ \left(3 - \frac{5q}{4} \right) J_1^2 J_2 - \left(1 - \frac{11}{4q} \right) J_1 J_2^2 \right]
 \cos(2 \theta_1 - 2 \theta_2) \label{K11-4}
\end{eqnarray}
and those of the generating function are
\begin{eqnarray}
G_{2} &=& 
- \frac{3 q}{32} P_{X}^3 X - \frac{3}{32} P_{X} P_{Y}^2 X 
- \frac{5 q}{32} P_{X}\ X^3 - \frac{3}{32} P_{X}^2\ P_{Y} Y - \nonumber\\
&& \frac{3}{32 q} P_{Y}^3 Y - \frac{5}{32} P_{Y} X^2 Y - 
    \frac{5}{32} P_{X} X Y^2 - \frac{5}{32 q} P_{Y} Y^3 , \label{g11-2} \\
G_{4} &=& \frac{5 q}{96} P_{X}^5 X - \frac{13 q^2}{256} P_{X}^5 X 
+ \frac{9}{128} P_{X}^3\ P_{Y}^2 X - \frac{13 q}{192} P_{X}^3\ P_{Y}^2 X  - \nonumber\\
&& \frac{13}{768} P_{X} P_{Y}^4 X + \frac{7}{384 q} P_{X} P_{Y}^4 X -
 \frac{19 }{768} P_{X} X Y^4 - \frac{19}{384}  P_{Y} X^2 Y^3  + \nonumber\\
&& \frac{9}{128}  P_{X} P_{Y}^2 X^3 - \frac{37 q}{384} P_{X} P_{Y}^2 X^3
+\frac{25}{192 q}  P_{Y} X^2 Y^3 - \frac{19 q^2}{256} P_{X} X^5  - \nonumber\\
&& \frac{9}{256} P_{X}^4 P_{Y} Y - \frac{13 q}{384} P_{X}^4 P_{Y} Y
- \frac{13}{384} P_{X}^2\ P_{Y}^3 Y + \frac{7}{192 q} P_{X}^2\ P_{Y}^3 Y + \nonumber\\
&& \frac{25 }{384 q} P_{X} X Y^4 + \frac{9}{64}  P_{X}^2 P_{Y} X^2 Y 
- \frac{5 q}{64} P_{X}^2 P_{Y} X^2 Y - \frac{5}{384} P_{Y}^3 X^2 Y  - \nonumber\\
&& \frac{5}{384 q} P_{Y}^3 X^2 Y + \frac{23}{256}  P_{Y} X^4 Y -
\frac{19 q}{384}  P_{Y} X^4 Y + \frac{9}{128} P_{X}^3 X Y^2 -\nonumber\\
&& \frac{37 q}{384} P_{X}^3 X Y^2 -\frac{7}{64} P_{X} P_{Y}^2 X Y^2 +
\frac{11}{64 q} P_{X} P_{Y}^2 X Y^2+ \frac{23}{128}  P_{X} X^3 Y^2 -\nonumber\\
&& \frac{19 q}{192} P_{X} X^3 Y^2 -\frac{5}{384} P_{X}^2 P_{Y} Y^3 -
\frac{5}{384 q} P_{X}^2 P_{Y} Y^3+ \frac{1}{288 q^2} P_{Y}^3 Y^3 +\nonumber\\
&& \frac{5 q}{36} P_{X}^3 X^3 - \frac{13 q^2}{96} P_{X}^3 X^3  + \frac{11 q}{96} P_{X} X^5 
 + \frac{31}{768 q^2} P_{Y} Y^5  + \frac{1}{768 q^2} P_{Y}^5 Y + \nonumber\\
&& \frac{\delta q}{32} \left(3 P_{X}^3 X + 5 P_{X} X^3   \right) +\nonumber\\
&& \frac{\delta}{64} \left( 7 P_{X} P_{Y}^2 X - P_{X}^2 P_{Y} Y + P_{Y} X^2 Y + 9 P_{X} X Y^2 \right). \label{g11-4}
\end{eqnarray}
For the 1:2 resonance the terms of the normal form are
\begin{eqnarray}
 K_{0} &=& J_1 + 2 J_2, \label{K12-0}\\
 K_{2} &=& 2 \delta J_1 - \frac34 \left(q J_{1}^2 + \frac1{q} J_{2}^2 \right) - J_{1} J_{2} , \label{K12-2}\\
 K_{4} &=& q \left(\frac56 - 
\frac{17}{16} q \right) J_{1}^3 + 
\frac{29}{96 q^{2}} J_{2}^3 + \nonumber\\ 
 && \left(\frac{13}{12} - 
\frac32 q \right) J_{1}^2 J_{2} - \left(\frac5{12} - 
\frac{3}{4q} \right) J_{1} J_{2}^2 + \nonumber\\ 
&& \frac18 q J_1^2 J_2 \cos(4 \theta_1 - 2 \theta_2) \label{K12-4}
\end{eqnarray}
and those of the generating function are
\begin{eqnarray}
G_{2} &=& 
- \frac{3 q}{16} P_{X}^3 X - \frac13 P_{X} P_{Y}^2 X 
- \frac{5 q}{16} P_{X}\ X^3 + \frac{1}{24} P_{X}^2\ P_{Y} Y - \nonumber\\
&& \frac{3}{32 q} P_{Y}^3 Y - \frac{7}{24} P_{Y} X^2 Y - 
    \frac16 P_{X} X Y^2 - \frac{5}{32 q} P_{Y} Y^3 , \label{g12-2} \\
G_{4} &=& \frac{5 q}{48} P_{X}^5 X - \frac{13 q^2}{64} P_{X}^5 X 
+ \frac{19}{144} P_{X}^3\ P_{Y}^2 X - \frac{53 q}{192} P_{X}^3\ P_{Y}^2 X  - \nonumber\\
&& \frac{4}{45} P_{X} P_{Y}^4 X + \frac{1}{32 q} P_{X} P_{Y}^4 X -
 \frac{13 }{90} P_{X} X Y^4 - \frac{17}{2880}  P_{Y} X^2 Y^3  + \nonumber\\
&& \frac{25}{144}  P_{X} P_{Y}^2 X^3 - \frac{67 q}{192} P_{X} P_{Y}^2 X^3
+\frac{65}{768 q}  P_{Y} X^2 Y^3 - \frac{19 q^2}{64} P_{X} X^5  - \nonumber\\
&& \frac{1}{144} P_{X}^4 P_{Y} Y - \frac{19 q}{384} P_{X}^4 P_{Y} Y
- \frac{67}{2880} P_{X}^2\ P_{Y}^3 Y + \frac{7}{256 q} P_{X}^2\ P_{Y}^3 Y + \nonumber\\
&& \frac{9 }{64 q} P_{X} X Y^4 + \frac{1}{24}  P_{X}^2 P_{Y} X^2 Y 
- \frac{5 q}{64} P_{X}^2 P_{Y} X^2 Y - \frac{83}{2880} P_{Y}^3 X^2 Y  - \nonumber\\
&& \frac{1}{256 q} P_{Y}^3 X^2 Y + \frac{17}{144}  P_{Y} X^4 Y -
\frac{q}{128}  P_{Y} X^4 Y + \frac{5}{36} P_{X}^3 X Y^2 -\nonumber\\
&& \frac{73 q}{192} P_{X}^3 X Y^2 -\frac{13}{60} P_{X} P_{Y}^2 X Y^2 +
\frac{7}{64 q} P_{X} P_{Y}^2 X Y^2+ \frac{19}{72}  P_{X} X^3 Y^2 -\nonumber\\
&& \frac{83 q}{192} P_{X} X^3 Y^2 -\frac{73}{2880} P_{X}^2 P_{Y} Y^3 +
\frac{25}{768 q} P_{X}^2 P_{Y} Y^3+ \frac{1}{288 q^2} P_{Y}^3 Y^3 +\nonumber\\
&& \frac{5 q}{18} P_{X}^3 X^3 - \frac{13 q^2}{24} P_{X}^3 X^3  + \frac{11 q}{48} P_{X} X^5 
 + \frac{31}{768 q^2} P_{Y} Y^5  + \frac{1}{768 q^2} P_{Y}^5 Y + \nonumber\\
&& \frac{\delta q}{8} \left(3 P_{X}^3 X + 5 P_{X} X^3   \right) +\nonumber\\
&& \frac{\delta}{9} \left( 2 P_{X} P_{Y}^2 X + 2 P_{X}^2 P_{Y} Y - 2 P_{Y} X^2 Y + 7 P_{X} X Y^2 \right). \label{g12-4}
\end{eqnarray}
Concerning these formulas, two remarks are in order:
\begin{enumerate}
      \item The monomials associated to the detuning (namely, with $\delta$ appearing in the coefficients) are of 2 degrees lower than that of the specific term of a given series. For example, in $G_{4}$ (degree 6) they appear with degree 4. This is due to the choice of considering the detuning term in (\ref{HS}) of 2nd order in the perturbation: it can be shown (Pucacco et al. \cite{PBB}) that this choice, in principle not unique, is the `optimal' one. Being present in $G$ at order 4, they appear in $K$ only at order 6.
         
      \item The normalizing variables $\vec{P},\vec{X}$ have to be considered as {\it new} canonical variables at each step of the normalization: so, for example, the $\vec{P},\vec{X}$ arguments of $G_{2}$ when truncating $G$ at $N=2$ are different from the arguments of $G_{2}$ when truncating $G$ at $N=4$. A notation able to represent these features could be introduced but it would be heavy and we prefer to stay with the standard practice of ignoring these subtleties. However, this observation gives reason for the apperance of the extra Poisson brackets in the transformations of the form (\ref{x5}).
         
   \end{enumerate}

\end{appendix}

\end{document}